\documentclass[jkps,preprint,fleqn,showpacs,showkeys]{revtex4}
\usepackage[pdftex]{graphicx}
\usepackage{amssymb}
\usepackage{amsmath}
\usepackage{bm}
\usepackage{amssymb}
\usepackage{amsmath}
\usepackage{txfonts}
\usepackage{mathdots}
\usepackage{graphicx}
\usepackage{subfigure}
\usepackage{bm}

\newcommand{\hbrn}{\hat{\rm{\bf n}}}
\newcommand{\hbru}{\hat{\rm{\bf u}}}
\newcommand{\hbrv}{\hat{\rm{\bf v}}}
\newcommand{\hbrx}{\hat{\rm{\bf x}}}
\newcommand{\hbry}{\hat{\rm{\bf y}}}
\newcommand{\hbrz}{\hat{\rm{\bf z}}}
\newcommand{\Y}{\Y_{lm}}

\newcommand{\be}{\begin{equation}}
\newcommand{\ee}{\end{equation}}
\newcommand{\Be}{\begin{eqnarray}}
\newcommand{\Ee}{\end{eqnarray}}

\newcommand{\f}{\frac}

\begin{document}
\title[]{Gravitational Waves by the Perturbation of a Rotating Axisymmetric Rigid Body}
\author{Sung-Won Kim}
\email[email:]{sungwon@ewha.ac.kr}
\affiliation{Ewha Womans University, Seoul 03760,
Korea}

\date{\today}

\begin{abstract}
Precession is one of the important mechanisms of gravitational wave generation in astrophysics.
In general, free precession of a rigid body can be caused by the rotation of a triaxial body. In the case of
symmetric body, if only the principal axis does not coincide with the axis of rotation, then there is the precession.
When a symmetric body rotates around one of its principal axes, the body cannot move with  precession.
However, when there is a perturbation in angular velocity of the symmetric body spinning around the principal axis of the largest moment, the body can have a precession motion.
In this paper, the wave forms and their characteristics of gravitational waves by the perturbation of a rotating axisymmetric rigid body are studied.
\end{abstract}

\maketitle

\section{Introduction}

For the gravitational wave generation by the motion of rotating rigid body,
the main motions affecting it are the precession and the nutation as well as rotation. The neutron stars and pulsars are good examples of the gravitational waves caused by precession motion of the bodies. Early treatment on gravitational waves by freely precessing neutron stars based on rigid body motion is conducted by Zimmermann et al \cite{ZZ}. Broeck studied the gravitational wave spectrum of non-axisymmetric freely precessing neutron stars \cite{BB} and Gao \& Shao derived the gravitational waves by precession of triaxially deformed neutron stars \cite{GS}.
There are also evidences for gravitational waves by free precession in the radio signature of PSR B1828-11 \cite{Evidence} and PSR B1642-03 \cite{SLU}.

As we know, free precession happens when triaxial ellipsoid rotates or the principal axis of an axisymmetric body does not coincide with the angular momentum \cite{Mechanics}.
If the axisymmetric body rotates around the principal axis, the body does not move with free precession.
However, the perturbation in angular velocity of the principal axis causes oscillation and precession of axis like the forced motion. In the astrophysical situations, there is a high probability of perturbation of the star due to the scattering of small bodies.

In this paper, we derive the gravitational waves generated by the perturbed angular velocity of a rotating axisymmetric rigid body, when it rotates around the principal axis.
In this case, the stable motion is under the condition that the principal moment around the rotating axis should be larger than the others.
The tip of the angular velocity vector has a two-dimensional harmonic motion in a plane \cite{Mechanics}.
For the perturbed case of an axisymmetric rigid body, the axis moves in circular, elliptic, or linear motions, according to the initial conditions.
In case of the linear motion, there is an oscillation of axis in a plane containing the symmetry principal axis.
When it moves in a circular motion, we see the precession of the principal axis.
If the axis moves in elliptical motion, it has the precession without wobbling because the tip moves in a plane.

Before processing the main issues of this paper, let's take a look at the gravitational waves caused by free precession of a rigid body in two cases: (1) triaxial body and (2) symmetric body rotating around a non-principal axis. The examples are well introduced in  references \cite{MM,CA}. Next, according to the method used in the unperturbed free precession cases, we derive the gravitational waves by the perturbation case of an axisymmetric body rotating around the principal axis.

\section{Basic Formulas}

Now we introduce the basic formulas on gravitational waves by quadrupole moment to derive the amplitudes of our cases following the book for gravitational waves \cite{MM}.
The gravitational waves are generated by the time dependent quadrupole moment of the body. The definitions of components of inertia tensor $\mathbb{I}$ and of the reduced form of quadrupole moment $\mathbb{Q}$ are as follows:
\Be
I_{ij} &=& \int (r^2 \delta_{ij} - x_ix_j) \rho dV, \label{inertia}\\
Q_{ij} &\equiv& M_{ij} - \f{1}{3}\delta_{ij}M_k^k = \int (x_ix_j - \f{1}{3}r^2 \delta_{ij} ) \rho dV,\label{quadrupole}
\Ee
where $\rho$ is the mass volume density of the body.
These two definitions (\ref{inertia}) and (\ref{quadrupole}) can be identified except the change of the sign, after the second derivation with respect to time, so that there is a relationship as $M_{ij} = -I_{ij} + c_{ij}$ with constants $c_{ij}$.
So, for rigid body, the inertia tensor is an important physical quantity that derives the frequency and amplitude of the gravitational waves generated by any motions of the body.

Because the gravitational wave is represented by the second derivatives of the transverse-traceless (TT) quadrupole moment, it can be represented by
\be
h_{ij}^{\mathrm{TT}}(t,r) = \f{2G}{rc^4} \ddot{Q}^{\mathrm{TT}}_{ij}(t-r/c).
\ee
We need the retarded time to see the gravitational wave from the body. Hereafter, the time used in $h$ means the retarded time $(t-r/c)$.
When we project the tensor on the $x$-$y$ plane or $y$-$z$ plane in deriving TT forms, respectively, two polarized wave forms are
\Be
h_+ &=& \f{G}{rc^4} (\ddot{M}_{11}-\ddot{M}_{22}) \quad \mbox{or} \quad \f{G}{rc^4} (\ddot{M}_{22}-\ddot{M}_{33}), \label{h+}\\
h_\times &=& \f{2G}{rc^4} \ddot{M}_{12} \quad \mbox{or} \quad \f{2G}{rc^4} \ddot{M}_{23},  \label{hx}
\Ee

\begin{figure}[t!]
  \centering
    \subfigure
    {%
    \includegraphics[width=0.4\textwidth]{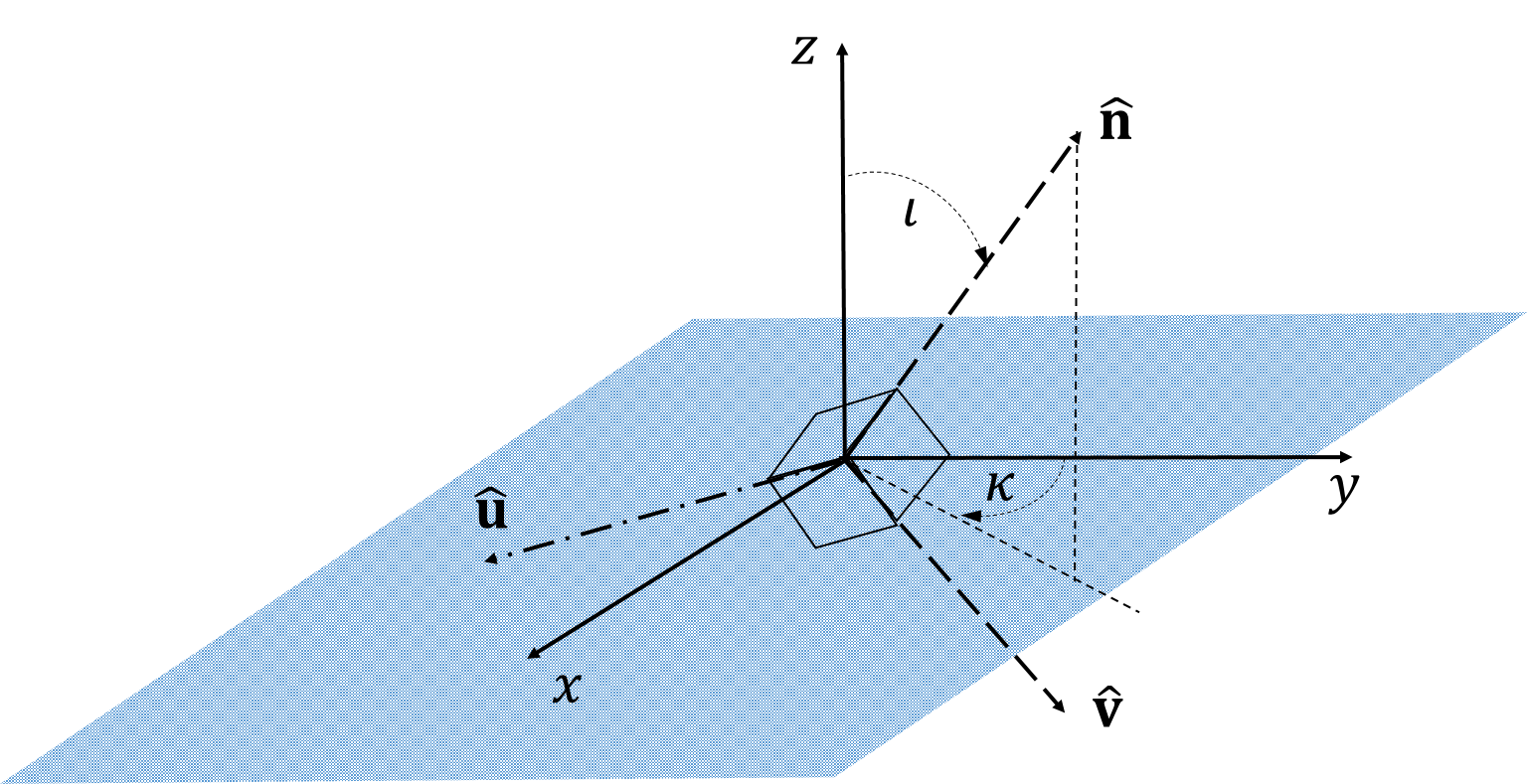}%
    }
  \caption{The relation between $(\hbrx,\hbry,\hbrz)$ frame and the $(\hbru,\hbrv,\hbrn)$ frame. The vector $\hbrn$ is at an angle $\iota$ from $z$-axis and $\kappa$ from $y$-axis \cite{MM}.}
  \end{figure}

To calculate the amplitudes for a gravitational waves propagating in the generic direction $\hbrn$, we introduce two unit vectors $\hbru$ and $\hbrv$ orthogonal to $\hbrn$ so that $\hbru \times \hbrv = \hbrn $ as shown in Fig.~1.
The values detected in the angle $\iota$ from the $z$-axis and the angle $\kappa$ from the $y$-axis are transformed value with rotation matrix
$\mathcal{R}_{\mathrm d}(\iota,\kappa)\equiv \mathcal{R}_z(\kappa)\mathcal{R}_x(\iota)$.
Here $R_z(\kappa)$ is the rotation matrix with angle $\kappa$ clockwise around $z$-axis such as
\[
\mathcal{R}_z(\kappa) = \left(
                \begin{array}{ccc}
                  \cos\kappa & \sin\kappa & 0 \\
                  -\sin\kappa & \cos\kappa & 0 \\
                  0 & 0 & 1 \\
                \end{array}
              \right).
\]
If we represent a tensor $\mathbb{M}$ in the $(x,y,z)$ frame as the tensor $\mathbb{M}'$ in the $(x',y',z')$ frame, whose axes are in the direction $(\hbru,\hbrv,\hbrn)$, we have the transformation between them as $\mathbb{M}' = \mathcal{R}_{\mathrm d}^{\mathrm T} \mathbb{M} \mathcal{R}_{\mathrm d}$, where $\mathcal{R}^{\mathrm T}$ is the transpose of the matrix $\mathcal{R}$.

The radiated energy is
\be
P = \f{1}{5} \f{G}{c^5} \langle \dddot{Q}_{ij}\dddot{Q}^{ij}\rangle =
\f{1}{5} \f{G}{c^5} \langle \dddot{M}_{ij}\dddot{M}^{ij} - \f{1}{3}(\dddot{M}_{kk})^2\rangle.
\ee
The angular momentum carried away, per unit time is
\be
\f{dJ^i}{dt}=\f{2G}{5c^5}\epsilon^{ikl} \langle {\ddot{Q}}_{ka}\dddot{Q}_l^a \rangle,
\ee
where $\epsilon^{ikl}$ is the Levi-Civita density. Therefore, the instantaneous rate of decrease of energy and angular momentum of the source are given by
\[
\f{dE_{\rm source}}{dt} = -P \quad \mbox{and} \quad \f{dL^i_{\rm source}}{dt} = - \f{dJ}{dt}.
\]

\section{Examples of unperturbed free precession}

\subsection{Triaxial Body}

\begin{figure}[t!]
  \centering
    \subfigure
    {%
    \includegraphics[width=0.3\textwidth]{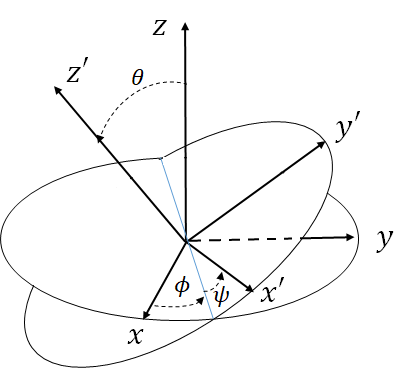}%
    }
  \caption{The coordinates in body frame $(x',y',z')$ and in fixed frame $(x,y,z)$ are related with Euler angles $(\phi,\theta,\psi)$.}
  \end{figure}

Let the principal moments of inertia of the rigid body $\mathbb{I}_\mathrm{b}$ be $I_1, I_2$, and $I_3$ in body frame $(x',y',z')$ as shown in Fig.~2.
As shown in textbook \cite{Mechanics}, the rotation of triaxial body around one of its principal axes will have a precession which gives rise to time dependence in the motion of the inertia tensor. 
The inertia tensor in the fixed frame is denoted by $\mathbb{I}_\mathrm{f}$.
Let the body rotate with the angular velocity $\omega$ around one of the principal axis which is $z$-axis both in body frame and in fixed frame.  Because $\mathbb{I}_{\mathrm b} = \mathcal{R}_z(\omega t)\mathbb{I}_{\mathrm f} \mathcal{R}_z(\omega t)^{\mathrm T}$, the transformation from inertia tensor in body frame to the inertia tensor in fixed frame, is done as
\[
\mathbb{I}_{\mathrm f} = \mathcal{R}_z(\omega t)^{\mathrm T} \mathbb{I}_{\mathrm b} \mathcal{R}_z(\omega t).
\]
Two polarized forms of the corresponding gravitational waves are, according to the formulas (\ref{h+}) and (\ref{hx}), given as \cite{CA}\cite{MM}
\Be
h_+(t) &=& \f{4G\epsilon_t I_3 \omega^2}{c^4r}\cos 2\omega t, \\
h_\times(t) &=&  \f{4G\epsilon_t I_3 \omega^2}{c^4r}\sin 2\omega t,
\Ee
where $\epsilon_t=(I_1-I_2)/I_3$ is known as the ellipticity of the triaxial body. The detection is achieved at the position $r$ from the gravitational source. The frequency of the gravitational waves is twice of the rotation frequency of the body. When the inclination angle is $\iota$ (see Fig.~1),
\Be
h_+(t;\iota) &=& \f{4G\epsilon_t I_3 \omega^2}{c^4r} \f{(1+\cos^2\iota)}{2} \cos 2\omega t,
\\
h_\times(t;\iota) &=&  \f{4G\epsilon_t I_3 \omega^2}{c^4r} (2\cos \iota) \sin 2\omega t,
\Ee
by transforming with $\mathcal{R}_{\mathrm{d}}(\iota,0)=\mathcal{R}_x(\iota)$.
In case of $\iota=\pi/2$, that is, the detection in the direction of perpendicular to $z$-axis and on the $x$-$y$ plane,
shows $h_+(t;\pi/2)=h_+(t)/2$ and $h_\times(t;\pi/2)=0$, only + polarization appear.
The total power radiated is
\be
P =
\f{32}{5} \f{G}{c^5}\epsilon_t^2 {I_3}^2 \omega^6.
\ee
The instantaneous rate of decreases of energy is $dE/dt = -P$.
The angular momentum lost is
\be
\f{dL_3}{dt} =  - \f{2}{5} \f{G}{c^5}\epsilon_{3ij}\langle \dddot{Q}^{ik}{\ddot{Q}^{j}}_k\rangle
= - \f{32}{5} \f{G}{c^5}\epsilon_t^2 {I_3}^2 \omega^5 =  \f{dE/dt}{\omega}.
\ee
Because the rotational energy of the body is $E=(1/2)I_3\omega^2$, the frequency decrease rate is given by
\be
\f{d\omega}{dt} = - \f{32}{5} \f{G}{c^5}\epsilon_t^2 {I_3} \omega^5.
\ee

\subsection{Symmetric Body}

For the case of symmetric body, for example $I_1 = I_2 < I_3$, such a precession of the triaxial body does not occur, since the ellipticity $\epsilon_t=0$.
Thus the symmetric body can generate gravitational wave by the more complicated motion.
When the rotation axis does not coincide with the principal axis of the body, the angular momentum rotates around the principal axis, that is the precession motion.
From the textbook on mechanics \cite{Mechanics}, by solving the force-free Euler equation we can see the motion of free precession as the principal axis rotates around the angular momentum axis with $\Omega_0$
\be
\Omega_0=\f{J}{I_1},
\ee
where the angular momentum is $\mathbf{J}=(0,0,J)$ in fixed frame.
In this case, the transformation of the inertia tensor rotating with $\Omega$ is required \cite{MM}.
We can get the values in fixed frame with the transformation of $\mathcal{R}_1 \equiv \mathcal{R}_x(\theta)\mathcal{R}_z(\Omega_0 t)$, if the angular momentum is $z$-axis in fixed frame. Similar to the triaxial case, $\mathbb{I}_{\mathrm f} = \mathcal{R}_1^{\mathrm T} \mathbb{I}_{\mathrm b} \mathcal{R}_1$,
the components of $\mathbb{I}$ and $\mathbb{M}$ in fixed frame are calculated.
When we detect it at an inclination angle $\iota$ from the $z$-axis (angular momentum vector), by using
the formulas (\ref{h+}) and (\ref{hx}) and $\mathcal{R}_x(\iota)$,
\Be
h_+(t;\iota,\kappa) &=& - \f{4G\epsilon_s I_1 \Omega_0^2}{c^4r}
[
(2\sin^2 \theta (1+\cos^2\iota) \cos 2\Omega_0 t 
+ (\sin 2\theta \sin\iota \cos \iota) \cos \Omega_0 t
] \\
h_\times(t;\iota,\kappa) &=&  - \f{4G\epsilon_s I_1 \Omega_0^2}{c^4r} [(4\sin^2\theta \cos\iota)\sin 2\Omega_0 t 
+ (\sin2\theta \sin\iota) \sin \Omega_0 t],
\Ee
where the new ellipticity is defined as
\[
 \epsilon_s = \f{I_3-I_1}{I_1}
\]
for the symmetrical oblate spheroidal body.  Here $\theta$ is the angle between $z$-axis and $z'$-axis, and it is time-independent. The resultant gravitational waves are mixing of two spectrums: $2\Omega_0$ and $\Omega_0$.
If the angle $\theta$ is very small, the $\Omega_0$ term is dominant comparing to the $2\Omega_0$ term.
When $\iota=0$, that is, the angular momentum direction, we detect the gravitational wave with the frequency $2\Omega_0$ only. The phase difference between two polarized forms $h_+$ and $h_\times$ is $\pi/2$. When $\iota=\pi/2$, $h_+$ is $2\Omega_0$ and $h_\times$ is $\Omega_0$.
The total power radiated is
\be
P = \f{2G}{5c^5}\epsilon_s^2 I_1^2 \Omega_0^6 \sin^2\theta (\cos^2\theta+16\sin^2\theta).
\ee
As we can see, the larger the difference of two principal components of inertia tensor, the larger the amplitude and radiated power. The angular momentum lost rate is
\be
\f{dL}{dt}=-\f{2G}{5c^5}\epsilon_s^2 I_1^2 \Omega_0^5 \sin^2\theta (\cos^2\theta+16\sin^2\theta)=\f{dE/dt}{\Omega_0}.
\ee
From this, we get
\be
\f{d{\Omega}_0}{dt} = -\f{2G}{5c^5}\epsilon_s^2 I_1 \Omega_0^5 \sin^2\theta (\cos^2\theta+16\sin^2\theta).
\ee

\section{Precession by Perturbation}

Now we investigate the gravitational waves by precession for the perturbed cases. 
The perturbation of the rotating symmetric axis also causes the precession motion of the body. 
Let the body rotate around $x_3$-axis with the angular velocity $\boldsymbol{\omega}=\omega_3\boldsymbol{e}_3$ and be exerted by a small perturbation.
The angular velocity vector assumes the form as
\[
\boldsymbol{\omega}=\lambda \boldsymbol{e}_1+ \mu {\boldsymbol{e}}_2+ \omega_3 \boldsymbol{e}_3,
\]
where $\lambda$ and $\mu$ are small quantities.
The Euler equations become
\Be
&&(I_2-I_3)\mu \omega_3 - I_1 \dot{\lambda} =0, \nonumber \\
&&(I_3-I_1)\lambda \omega_3 - I_2 \dot{\mu} =0, \nonumber \\
&& (I_1-I_2)\lambda \mu - I_3 \dot{\omega}_3 =0.
\Ee
If the body is the axisymmetric body, namely, $I_1 =I_2$, the solutions to the Euler equations are $\omega_3$ = const and
\Be
\omega_1 &=& \lambda(t)= A \cos(\Omega t-\alpha),\label{per:w1}\\
\omega_2 &=& \mu(t) = B \cos(\Omega t-\beta),\label{per:w2}
\Ee
where
\[
\Omega=\omega_3\sqrt{\frac{(I_3-I_1)(I_3-I_2)}{I_1I_2}}=\omega_3\epsilon_s.
\]
Thus, the perturbation in angular velocity causes a two-dimensional harmonic motion of the two components $\omega_1$ and $\omega_2$ of the angular velocity.
There are three motions dependent on the initial conditions of perturbation.

\begin{figure}[t!]
  \centering
    \subfigure
    {%
    \includegraphics[width=0.3\textwidth]{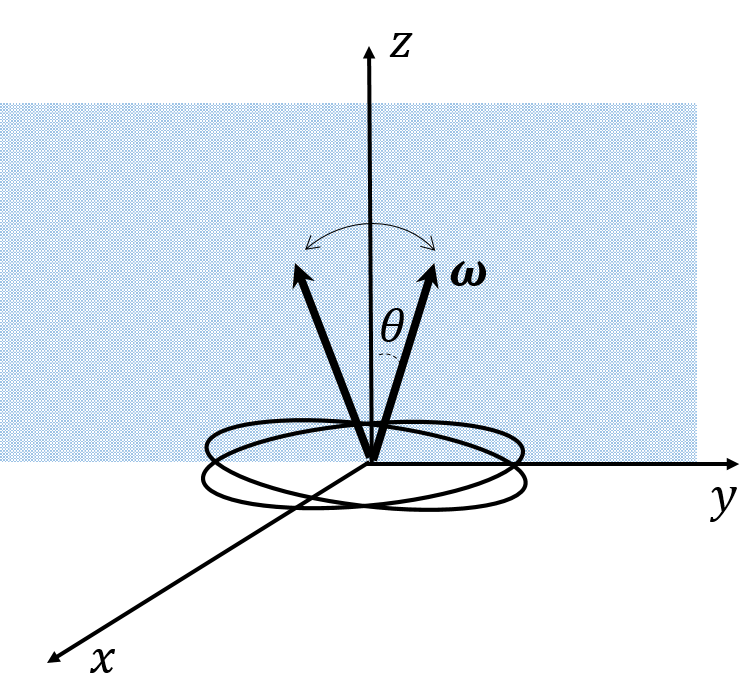}%
    }
  \caption{The linear oscillation by perturbation. The angular velocity vector $\boldsymbol{\omega}$ oscillates in $yz$-plane and the tip of the axis moves linearly in a plane parallel to $xy$-plane.}
  \end{figure}

\subsection{Linear Oscillation}

If the phase difference of two resultant motion $\alpha-\beta$   in (\ref{per:w1}) and (\ref{per:w2}) is 0, the motion of the tip of the rotation axis is the linear oscillatory motion with the frequency $\Omega$ in plane including $z$-axis in fixed frame as we see in Fig.~3.
We can consider this linear motion as the partial rotation around an axis on $x$-$y$ plane.
If we set $A=B$ and the phase constants $\alpha=\beta=0$ so that  $\omega_1=\omega_2=A\cos\Omega t$ which is maximum at $t=0$.
If we rotate clockwise $x$- and $y$-axes with the angle $\pi/4$  around $z$-axis, the angular velocity becomes
\[
\boldsymbol{\omega}= \omega'_1 \boldsymbol{e}'_1 + \omega_3 \boldsymbol{e}_3,
\]
with new unit vector $\boldsymbol{e}'_1$ along the new $x$-axis.
We can consider the perturbation $\omega'_1=\sqrt{2}A\cos\Omega t$ of the body
as the oscillatory rotation with an angle $\theta$ around $x$-axis (See Fig.~2).
The transformed inertia tensor becomes
\be
\mathbb{I}_{\mathrm{f}}
=
 \mathcal{R}_x(\theta)^\mathrm{T}\mathcal{R}_z({\scriptstyle{\f{\pi}{4}}})^\mathrm{T} \mathbb{I}_\mathrm{{b}} \mathcal{R}_z({\scriptstyle{\f{\pi}{4}}})\mathcal{R}_x(\theta) =\mathcal{R}_x(\theta)^\mathrm{T} \mathbb{I}_\mathrm{{b}} \mathcal{R}_x(\theta)
=
\frac{I_1}{2}\left( \begin{array}{ccc}
2 & 0 & 0 \\
0 & 2+\epsilon (1-\cos2\theta ) & \epsilon \sin 2\theta  \\
0 & \epsilon \sin 2\theta & 2+\epsilon (1+\cos 2\theta ) \end{array}
\right).
\ee
The $\theta$ is derived from $\omega_1'$ as
\[
\theta = \int \omega_1' dt = \eta \sin\Omega t + \theta_c.
\]
By setting integration constant $\theta_c$ to zero, the body initially
rotates without axis tilt, i.e. $\theta(0)=0$.
The amplitude of $\theta$ is $\eta \equiv \sqrt{2}A/\Omega$ which is equal to
the ratio of the perturbation size to ellipticity, $\sqrt{2}\Delta/\epsilon_s$, where $\Delta\equiv A/\omega_3$ is the
fractional perturbation.
Its time derivatives are
\[
\dot{\theta} =  \eta \Omega \cos\Omega t \quad \mbox{and}
\quad \ddot{\theta}=-\eta \Omega^2 \sin\Omega t.
\]
If we take the double derivative to the inertia tensor, then we get
\be
\ddot{M}(t)
=
I_1\epsilon_s \ddot{\theta}
\left( \begin{array}{ccc}
0 & 0 & 0 \\
0 & \sin 2\theta & -\cos 2\theta \\
0 & -\cos 2\theta & -\sin 2\theta
 \end{array}
\right) 
~+2I_1\epsilon_s \dot{\theta}^2
\left( \begin{array}{ccc}
0 & 0 & 0 \\
0 & -\cos 2\theta & -\sin 2\theta \\
0 & -\sin 2\theta & \cos 2\theta
 \end{array}
\right). \label{M_linear}
\ee
This is already trace-free and transverse with respect to $x$-axis.
Thus we get two polarized amplitudes of strain are
\Be
h_+(t)
&=& h_1 \left[ \sin \Omega t \sin \left( 2\eta \sin\Omega t \right)
- 2\eta\cos^2\Omega t \cos\left( 2\eta \sin\Omega t \right)\right], \label{nu_h+}\\
h_\times(t)
&=& - h_1 \left[ \sin \Omega t \cos\left( 2\eta \sin\Omega t \right)
+ 2\eta\cos^2\Omega t \sin\left( 2\eta \sin\Omega t \right)\right], \label{nu_hx}
\Ee
where
\[
h_1=\f{2G}{rc^4}\eta I_1\epsilon_s \Omega^2.
\]
The gravitational wave frequency $\Omega$ is given by $\Omega=\epsilon_s\omega_3$.
There might be an ambiguity of the size $\eta$, since $\epsilon_s$ is the characteristics of the oblate spheroid and cannot be arbitrarily small. However, if $\eta$ is very small, that is, the perturbation is small compared to $\Omega$,
\Be
h_+(t) & \simeq & - 2 h_1 \eta \cos 2\Omega t, \\
h_\times(t) & \simeq &  - h_1 \left[1 + 2\eta^2\left(2\cos^2\Omega t - \sin^2\Omega t \right)\right]\sin \Omega t.
\Ee


Let the observing direction be the inclination angle $\iota$ from $z$-axis and the angle $\kappa $ from y-axis, as seen in Fig.~1.
Two polarizations of radiation in $\hbrn$-direction are
\Be
h_+(t;\iota,\kappa) &=& \f{G}{rc^4} [ \ddot{M}_{22}(\cos^2\kappa +1)
(2\cos^2 \iota -1 ) -2\ddot{M}_{23} \cos\kappa \sin 2\iota ],
\nonumber \\
h_\times(t;\iota,\kappa) &=& \f{2G}{rc^4} [ \ddot{M}_{22}(\cos^2\kappa +1)
\sin \iota \cos \iota + \ddot{M}_{23} \cos\kappa \cos 2\iota ].
\nonumber
\Ee
If $\kappa =0$, the detection at the inclination angle $\iota $ from $z$-axis and along the $y$-axis, which is perturbation direction, two polarized amplitudes are
\Be
h_+(t;\iota,0) &=& \f{2G}{rc^4} [ \ddot{M}_{22} \cos 2\iota - \ddot{M}_{23} \sin 2\iota ], \label{h+0}\\
h_\times(t;\iota,0) &=& \f{2G}{rc^4} [ \ddot{M}_{22} \sin 2\iota + \ddot{M}_{23} \cos 2\iota ].  \label{hx0}
\Ee
In other words, the physical meaning of the two amplitudes, (\ref{h+0}) and (\ref{hx0}) can be interpreted as the whole
quantity by addition of two angles $\theta$ and $\iota$ like
\Be
h_+ &=& h_1  \sin2(\theta+\iota)\sin \Omega t, \nonumber \\
h_\times &=& h_1 \cos2(\theta+\iota) \sin\Omega t, \nonumber
\Ee
under the assumption that the second term in (\ref{M_linear}) is neglected.
When $\iota=0$, they are the same as (\ref{nu_h+}) and (\ref{nu_hx}) respectively,
and the negative signs are added when $\iota=\pi/2$.
And when $\iota=\pi/4$, two amplitudes are reversed with negative sign to $h_+$.

If $\kappa =\pi /2$, inclination angle $\iota $ from $z$-axis and along the $x$-axis which is perpendicular to the plane of perturbation, two polarized strains are
\Be
h_+(t;\iota,\pi/2) &=&  \f{G}{rc^4} \ddot{M}_{22}\cos 2\iota, \\
h_\times(t;\iota,\pi/2) &=& \f{G}{rc^4} \ddot{M}_{22}\sin 2\iota.
\Ee
They are same order of magnitude. When $\iota = 0$, $h_+$ and $h_\times$ are all zero, and we see $h_+=0, h_\times \neq 0$ for $\iota=\pi/4$, Also $h_+ \neq 0, h_\times =0$ when $\iota=\pi/2$.

The power radiated is
\[
P
=
\f{G}{5c^5} I_1^2 \epsilon_s^2 \eta^2 \Omega^6\left(1+\eta^2 \f{137}{16} \right)
\simeq
\f{G}{5c^5} I_1^2 \epsilon_s^2 \eta^2 \Omega^6 =\f{2G}{5c^5} I_1^2 \Delta^2 \Omega^6 
\]
with the dominant term.
The angular momentum by perturbation around $x$-axis lost rate is
\[
\f{dL_1}{dt}=-\f{4G}{5c^5} (I_1\epsilon_sA)^2\Omega^3 \langle \ddot{M}_{22}\dddot{M}_{32} + \ddot{M}_{23} \dddot{M}_{33} \rangle = 0.
\]
The time averages of each term are vanished and so we can say that the averaged angular momentum is conserved for this system.


\subsection{Circular Precession}

\begin{figure}[t!]
  \centering
    \subfigure
    {%
    \includegraphics[width=0.3\textwidth]{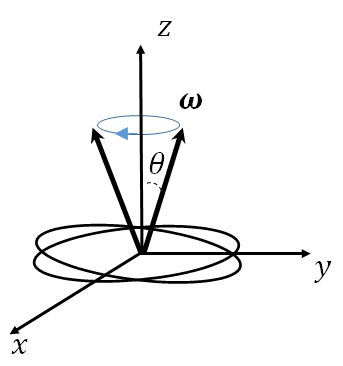}%
    }
  \caption{Circular and elliptic precessions around $z$-axis. The tip of $\boldsymbol{\omega}$ moves in a circle and an ellipse on the plane parallel to $x$-$y$ plane.   The former case has time-independent angle $\theta$ from $z$-axis and the latter has time-dependent $\theta$.}
\end{figure}

If $A=B$ and the phase difference is $\alpha-\beta=\pi/2$, it is a circular motion around $x_3$-axis in fixed frame, which is precession motion
and the angle $\theta=\theta_0$ is constant. See the Fig.~4. Thus we need the transformation
$\mathcal{R}_2\equiv\mathcal{R}_x(\theta_0)\mathcal{R}_z(\Omega t)$ from $\mathbb{I}_\mathrm{b}$
to $\mathbb{I}_\mathrm{f}$ and
we get the consequent components of $\mathbb{I}_\mathrm{f}$ as
\Be
I_{11} &=& I_1 (\cos^2\Omega t + \cos^2\theta_0 \sin^2\Omega t)+I_3 \sin^2\theta_0 \sin^2\Omega t, \nonumber\\
I_{12} &=&  (I_1-I_3)\sin^2\theta_0 \sin \Omega t \cos \Omega t, \nonumber \\
I_{13} &=& -(I_1-I_3)\sin\theta_0 \cos \theta_0 \sin \Omega t, \nonumber \\
I_{21} &=& (I_1-I_3)\sin^2\theta_0 \sin \Omega t \cos \Omega t, \nonumber \\
I_{22} &=&  I_1(\sin^2\Omega t+ \cos^2\theta_0 \cos^2\Omega t)+I_3 \sin^2\theta_0 \cos^2\Omega t, \nonumber\\
I_{23} &=& (I_1-I_3)\sin^2\theta_0 \sin \Omega t \cos \Omega t, \nonumber \\
I_{31} &=& -(I_1-I_3)\sin\theta_0 \cos \theta_0 \sin \Omega t, \nonumber \\
I_{32} &=&  (I_1-I_3)\sin\theta_0 \cos \theta_0 \cos \Omega t, \nonumber \\
I_{33} &=&  I_1 \sin^2\theta_0+I_3\cos^2 \theta_0. \nonumber 
\Ee

The second derivatives of the inertia tensor with respect to time are
\be
\ddot{M}(t) = - \f{1}{2} \Omega^2 (I_3-I_1) 
\left(
  \begin{array}{ccc}
   4\sin^2\theta_0 \cos 2\Omega t & 4\sin^2\theta_0 \sin 2\Omega t  &  -\sin 2\theta_0 \sin \Omega t \\
    4\sin^2\theta_0 \sin 2\Omega t & -4\sin^2\theta_0 \cos 2\Omega t &  \sin 2\theta_0 \cos \Omega t \\
   -\sin 2\theta_0  \sin \Omega t & \sin 2\theta_0  \cos \Omega t & 0 
  \end{array}
\right). \label{M0} 
\ee
Therefore the strains $h_+$ and $h_\times$ in direction $\hbrn$ with inclination angle $\iota$ are
\Be
h_+(t;\iota) &=& \f{G}{rc^4}( \ddot{M}_{11}-\ddot{M}_{22}\cos^2\iota - \ddot{M}_{33}\sin^2 \iota +
\ddot{M}_{23}\sin 2\iota \nonumber \\
&\simeq& h_0 [2 \Delta^2 (1+\cos^2\iota)\cos 2\Omega t + 2\Delta \sin\iota \cos\iota \cos \Omega t],
\\
h_\times(t;\iota)
&=& \f{G}{rc^4}( 2\ddot{M}_{12}\cos\iota - 2\ddot{M}_{13}\sin\iota) \nonumber \\
&=& h_0 ( 4\Delta^2 \cos\iota \sin 2\Omega t + 2\Delta \sin\iota \sin\Omega t ),
\Ee
where
\[
h_0 = -\f{G}{rc^4}I_1\epsilon_s\Omega^2.
\]
The results are the same as the previous free precession case of symmetric body except that
$\theta_0 = A/\omega_3=\Delta$ and $\Omega=\epsilon_s\omega_3$.
When $\iota=0$,
\[
h_+(t;0) = 4h_0 \Delta^2 \cos 2\Omega t, \qquad  h_\times(t;0) = 4h_0 \Delta^2 \sin 2\Omega t.
\]
They are same spectrum of previous cases and the order of magnitude is $\epsilon_s^3\Delta^2$.
When $\iota=\pi/2$,
\[
h_+(t;\pi/2) = 2h_0 \Delta^2 \cos 2\Omega t, \qquad  h_\times(t;\pi/2) = 2h_0 \Delta \sin \Omega t.
\]
In this case, the size of $h_+(t;\pi/2)$ is smaller than one of $h_\times$ in $\Delta$, and is half of $h_+(t;0)$. Note that only the $\Omega$ spectrum, not $2\Omega$ one, appears in $h_\times(t;\pi/2)$.
The power radiated is
\Be
P &=& \f{2G}{5c^5}I_1^2\epsilon_s^2\Omega^6 \sin^2\theta_0 (\cos^2\theta_0+16\sin^2\theta_0) \nonumber \\
 &\simeq & \f{2G}{5c^5}I_1^2\epsilon_s^8\omega_3^6 \Delta^2.  \label{RadPrec}
\Ee
for small angle $\theta_0$. It depends on $\epsilon_s^8$ and $\Delta^2$.
The angular momentum change rate is
\Be
\f{dL}{dt} &=& - \f{2G}{5c^5}I_1^2\epsilon_s^2\Omega^5 \sin^2\theta_0 (\cos^2\theta_0+16\sin^2\theta_0) \nonumber \\
 &\simeq & - \f{2G}{5c^5}I_1^2\epsilon_s^7\omega_3^5 \Delta^2. \label{J_Prec}
\Ee
for small angle $\theta_0$. Therefore
\[
\f{dJ}{dt}=\f{dE}{dt}\f{1}{\Omega} = \f{dE}{dt}\f{1}{\epsilon_s\omega_3}.
\]

\subsection{Elliptic Precession}

If $A\neq B$ and phase difference $\alpha-\beta=\pi/2$, the rotation axis moves forming an ellipse around $z$-axis which is the precession motion also. The inclination angle $\theta$ is not constant, but the tip of the axis is still in a plane so that there is no wobbling motion. When $\theta$ is very small,
\Be
\theta &\cong& \tan\theta = \f{1}{\omega_3}\sqrt{A^2\cos^2\Omega t + B^2\sin^2 \Omega t}
\nonumber \\
&=&\f{A}{\omega_3}\sqrt{\cos^2\Omega t + k^2\sin^2 \Omega t}   \cong \Delta \left( 1+ \f{\delta}{2}\sin^2 \Omega t \right), \label{theta_t}
\Ee
where $k^2 = B^2/A^2 = 1+ \delta$ and $\delta$ is considered as the small quantity. Here $\delta$ has the physical meaning of eccentricity $e^2$ of the ellipse,
so they have the relationship as
\[
e^2 = \left\{ \begin{array}{ccc}
               \delta  & \mbox{for} & B>A \\
                - \delta & \mbox{for} & A>B
              \end{array}
             \right. .\]
The time derivatives of $\theta$ are
\[
\dot{\theta}=\Delta \frac{\delta}{2}\Omega \sin 2\Omega t \quad \mbox{and} \quad \ddot{\theta}=\Delta\delta \Omega^2\cos 2\Omega t.
\]
To find the inertia tensor in fixed frame, we need the elliptic rotation transformation instead
of the circular one. The matrix that operates the elliptic rotation angle $\phi$ around $z$-axis is given as \cite{MO}
\[
\mathcal{E}_z(\phi)=
\left( \begin{array}{ccc}
\cos \phi  & \frac{1}{k}\sin \phi  & 0 \\
-k\sin \phi & \cos \phi  & 0 \\
0 & 0 & 1 \end{array}
\right),
\]
which is not orthogonal in general.
However, $\det|\mathcal{E}_z|=1$ and $\mathcal{E}_z^{\mathrm T}$ is used for transformation of the inertia tensor. The inertia tensor in fixed frame becomes
$
\mathbb{I}_{\mathrm f}=\mathcal{R}_3^{\mathrm T}\mathbb{I}_{\mathrm b}\mathcal{R}_3,
$
where
\[
\mathcal{R}_3\equiv\mathcal{R}_x(\theta)\mathcal{E}_z(\Omega t) =
\left( \begin{array}{ccc}
\cos\Omega t & \frac{1}{k}\sin\Omega t & 0 \\
-k \cos \theta \sin \Omega t & \cos \theta \cos \Omega t & \sin \theta  \\
k\sin \theta \sin \Omega t & -\sin \theta \cos \Omega t & \cos \theta  \end{array}
\right).
\]

Its motion is the rotation around the principal axis ($z'$-axis) with precession around $z$-axis. Neglecting $O(\delta^2)$ with the definition of $\theta$ (\ref{theta_t}),  we got the second derivatives of $\mathbb{M}$ given by
\Be
\ddot{M}_{11}
&=& -2(I_3-I_1)\Delta^2 \Omega^2 \left[  \cos 2\Omega t + \delta (\cos 2\Omega t + \sin^2 2\Omega t + 2 \sin^2 \Omega t \cos \Omega t) \right] - 2I_1 \delta \Omega^2 \cos 2\Omega t,
\nonumber\\
\\
\ddot{M}_{22}
&=& 2(I_3-I_1) \Delta^2 \Omega^2 (\cos 2\Omega t - \delta \cos 4\Omega t ) + 2 I_1 \delta \Omega^2 \cos 2\Omega t, \\
\ddot{M}_{33}
&=&  2(I_3-I_1)\Delta^2 \delta \Omega^2 \cos 2\Omega t, \\
\ddot{M}_{12}
&=& -\f{1}{2}(I_3-I_1) \Delta^2 \Omega^2 \left[ 4 \sin 2\Omega t + \delta (2\sin 2\Omega t - 3\sin 4\Omega t + 4 \sin^2 \Omega t \sin 2 \Omega t ) \right] \nonumber \\
&& ~~~- 2 I_1 \delta \Omega^2 \sin 2\Omega t, \\
\ddot{M}_{13}
&=& (I_3-I_1)\Delta \Omega^2 \left[ \sin \Omega t + \f{1}{2}\delta ( \sin \Omega t - 3 \sin 2\Omega t \cos \Omega t  + 3 \sin^2 \Omega t ) \right],\\
\ddot{M}_{23}
&=& - (I_3-I_1) \Delta \Omega^2 \left[ \cos  t + \delta \left( - \cos 3\Omega t + \f{1}{2}\sin^2 \Omega t \cos \Omega t \right) \right] .
\Ee
If $\delta=0$, then $\theta$ is constant and they are the same as (\ref{M0}). If we want to see the precession effect, we need the projection on $x$-$y$ plane, for transverse action.
we need $\ddot{M}_{11}, \ddot{M}_{22}$ and $\ddot{M}_{12}$. Therefore,
\Be
h_+ &=& - \f{G}{rc^4}\left\{(I_3-I_1)\Omega^2\Delta^2 (4\cos 2\Omega t + 2\delta [\cos 2\Omega t + \sin^2 2\Omega t + 2\sin^2\Omega t \cos 2\Omega t - \cos 4\Omega t ] ) \right.
\nonumber \\
&&\left.
-4\delta I_1\Omega^2  \cos 2\Omega t \right\},
\label{h+_ellipse} \\
h_\times
&=&
- \f{G}{rc^4}\left\{ (I_3-I_1)\Omega^2 \Delta^2 (4\sin\Omega t +
\delta [ 2 \sin 2\Omega t - 3 \sin 4\Omega t + 4\sin^2\Omega t \sin 2\Omega t ]) \right.
\nonumber \\
&&\left.
+ 2\delta \Omega^2 I_1 \sin 2\Omega t\right\}. \label{hx_ellipse}
\Ee
They are the same as the case of circular precession when $\delta=\dot{\theta}=0$.
There are three kinds of perturbed terms: the ellipticity of the inertia tensor $\epsilon_s$ or $\Omega$, the fractional perturbation of angular velocity $\Delta=A/\omega$, and the eccentricity of the trajectory of the principal axis $\delta$.

In the power radiated, there are four kinds of combinations from two small quantities $\Delta$ and $\delta$. The calculated terms are $\Delta^2, \Delta^2 \delta, \Delta^4,$ and $\Delta^4\delta$ such as
\[
P 
=
\f{2G}{5c^5}\epsilon_s^2\Omega^6 [ (I_3-I_1)^2\Delta^2  (1+\delta) + 32(I_3-I_1)I_1\Delta^2\delta 
+ (I_3-I_1)^2\Delta^4 (16+32\delta)]. 
\]
If we take the two leading terms,
\Be
P &\simeq & \f{2G}{5c^5}\Omega^6\Delta^2[(I_3-I_1)^2(1+\delta)+32(I_3-I_1)I_1\delta] \nonumber \\
&=& \f{2G}{5c^5}\Delta^2I_1^2[\epsilon_s^8 (1+\delta)+32\epsilon_s^7\delta]
\Ee
Of course, it is the same value as (\ref{RadPrec}), when $\delta=0$.
The angular momentum lost rate is
\[
\f{dL_3}{dt} 
\simeq
- \f{2G}{5c^3}\Omega^5\Delta^2[(I_3-I_1)^2(1+\delta)+32(I_3-I_1)I_1\delta]
=
\f{1}{\Omega}\f{dE}{dt}. 
\]

\section{Conclusion}

In this paper, we showed the characteristics of the gravitational waves for the perturbation of the symmetric rigid body rotating with a principal axis.
When the rotating symmetric body is perturbed in angular velocity vector, the tip of vector moves two dimensional harmonic motion in a plane, if the principal moment around the symmetric axis is larger than other moments.

Thus we exploited the gravitational waves by the three motions of the rotating axisymmetric body: linear motion, circular motion, and elliptic motion, according to the initial condition of the perturbation in angular velocity. The frequency of the gravitational waves by the perturbation is the ellipticity times the component of the angular velocity projected onto the rotation axis.

Table 1 shows the angular frequency and power radiated of the gravitational waves in each case. The power is represented as the dependencies of $\epsilon_t, \epsilon_s, \Delta$, and $\delta$, in order to see how small it is.
Among them, the ellipticity is not generally considered a small quantity,
and the powers can be roughly compared in magnitude of smallness with $\Delta$ and $\delta$. Of course, if $\epsilon$ of the system is very small, it also will play the role of measuring the smallness.

In perturbation case, the size of power radiated is proportional to the square of the fractional perturbation times $\epsilon_s^6$ for linear motion. 
The second case shows the circular motion which makes the circular precession whose amplitudes are the same as one of the unperturbed free precession except that ellipticity-dependent frequency, so that their amplitudes are
proportional to the square of the fractional perturbation times $\epsilon_s^8$.
When the tip of the angular velocity vector moves elliptic motion, the rigid body moves with elliptic precession without wobbling motion, though the inclination angle is time-dependent. The amplitude has the eccentricity correction terms comparing to circular motion.
These results can be applied to various astrophysical objects, especially if they rotate around a principal axis in a disturbing situation, such as the scattering by small objects.

\begin{table}[t!]
\caption{Smallness of frequency and power radiated of gravitational waves}
\begin{tabular}{cccccc}
  \hline
  & \multicolumn{2}{c}{~unperturbed free precession~} & \multicolumn{3}{c}{~~precession by perturbation ($\Delta$)~~ } \\\cline{2-6}
 & ~triaxial~ & ~symmetric~ & ~~linear~~ & ~circular~ & ~elliptic ($\delta$)~ \\
\hline
angular frequency $\Omega_{grav}$ & $2\omega$ & $2\Omega_0$ & $2\omega_3\epsilon_s$ & 2$\omega_3\epsilon_s$ & 2$\omega_3\epsilon_s$ \\ \hline
power radiated $P_{rad}$  & $\epsilon_t^2$ & $\epsilon_s^2$ & $\epsilon_s^6\Delta^2$ & $\epsilon_s^8\Delta^2$ &$\epsilon_s^7\Delta^2(\epsilon_s+\delta) $ \\
   \hline
\end{tabular}
\end{table}

\begin{acknowledgments}
This work was supported by National Research Foundation of Korea (NRF) funded by the Ministry of Education (NRF-2021R1I1A1A01056433).
\end{acknowledgments}

\end{document}